\documentclass[aps,prl,twocolumn,longbibliography,showpacs,amsmath,amsmath,amssymb,amsfonts,superscriptaddress]{revtex4-1}
\usepackage{graphicx}
\usepackage{dcolumn}
\usepackage{bm}
\usepackage{color}
\usepackage{multirow}
\usepackage{color}

\begin{document}
\title{Experimental signatures of an absorbing-state phase transition in an open driven many-body quantum system}

\author{Ricardo Guti\'errez}
\affiliation{School of Physics and Astronomy, University of Nottingham, Nottingham, NG7 2RD, UK}
\affiliation{Centre for the Mathematics and Theoretical Physics of Quantum Non-Equilibrium Systems, University of Nottingham, Nottingham, NG7 2RD, UK}
\affiliation{Complex Systems Group, Universidad Rey Juan Carlos, 28933 M\'{o}stoles, Madrid, Spain}
\author{Cristiano Simonelli}
\affiliation{INO-CNR, Via G. Moruzzi 1, 56124 Pisa, Italy}
\affiliation{Dipartimento di Fisica ``E. Fermi'', Universit\`a di Pisa, Largo Bruno Pontecorvo 3, 56127 Pisa, Italy}
\author{Matteo Archimi}
\affiliation{Dipartimento di Fisica ``E. Fermi'', Universit\`a di Pisa, Largo Bruno Pontecorvo 3, 56127 Pisa, Italy}
\author{Francesco Castellucci}
\affiliation{Dipartimento di Fisica ``E. Fermi'', Universit\`a di Pisa, Largo Bruno Pontecorvo 3, 56127 Pisa, Italy}
\author{Ennio Arimondo}
\affiliation{INO-CNR, Via G. Moruzzi 1, 56124 Pisa, Italy}
\affiliation{Dipartimento di Fisica ``E. Fermi'', Universit\`a di Pisa, Largo Bruno Pontecorvo 3, 56127 Pisa, Italy}
\affiliation{CNISM UdR Dipartimento di Fisica ``E. Fermi'', Universit\`a di Pisa, Largo Bruno Pontecorvo 3, 56127 Pisa, Italy}
\author{Donatella Ciampini}
\affiliation{INO-CNR, Via G. Moruzzi 1, 56124 Pisa, Italy}
\affiliation{Dipartimento di Fisica ``E. Fermi'', Universit\`a di Pisa, Largo Bruno Pontecorvo 3, 56127 Pisa, Italy}
\affiliation{CNISM UdR Dipartimento di Fisica ``E. Fermi'', Universit\`a di Pisa, Largo Bruno Pontecorvo 3, 56127 Pisa, Italy}
\author{Matteo Marcuzzi}
\affiliation{School of Physics and Astronomy, University of Nottingham, Nottingham, NG7 2RD, UK}
\affiliation{Centre for the Mathematics and Theoretical Physics of Quantum Non-Equilibrium Systems, University of Nottingham, Nottingham, NG7 2RD, UK}
\author{Igor Lesanovsky}
\affiliation{School of Physics and Astronomy, University of Nottingham, Nottingham, NG7 2RD, UK}
\affiliation{Centre for the Mathematics and Theoretical Physics of Quantum Non-Equilibrium Systems, University of Nottingham, Nottingham, NG7 2RD, UK}
\author{Oliver Morsch}
\affiliation{INO-CNR, Via G. Moruzzi 1, 56124 Pisa, Italy}
\affiliation{Dipartimento di Fisica ``E. Fermi'', Universit\`a di Pisa, Largo Bruno Pontecorvo 3, 56127 Pisa, Italy}

\newcommand{\changer}[1]{\textcolor{magenta}{#1}}

\keywords{}
\begin{abstract}
Understanding and probing phase transitions in non-equilibrium systems is an ongoing challenge in physics. A particular instance are phase transitions that occur between a non-fluctuating absorbing phase, e.g., an extinct population, and one in which the relevant order parameter, such as the population density, assumes a finite value. Here we report the observation of signatures of such a non-equilibrium phase transition in an open driven quantum system. In our experiment rubidium atoms in a quasi one-dimensional cold disordered gas are laser-excited to Rydberg states under so-called facilitation conditions. This conditional excitation process competes with spontaneous decay and leads to a crossover between a stationary state with no excitations and one with a finite number of excitations. We relate the underlying physics to that of an absorbing state phase transition in the presence of a field (i.e. off-resonant excitation processes) which slightly offsets the system from criticality. We observe a characteristic power-law scaling of the Rydberg excitation density as well as increased fluctuations close to the transition point. Furthermore, we argue that the observed transition relies on the presence of atomic motion which introduces annealed disorder into the system and enables the formation of long-ranged correlations. Our study paves the road for future investigations into the largely unexplored physics of non-equilibrium phase transitions in open many-body quantum systems.
\end{abstract}


\maketitle

Absorbing state phase transitions are among the simplest non-equilibrium phenomena displaying critical behavior and universality. They can occur for instance in models describing the growth of bacterial colonies or the spreading of an infectious disease among a population (see, e.g., \cite{Grassberger1983, Kuhr2011, Bonachela2012}). Once an absorbing state, e.g., a state in which all the bacteria are dead, is reached, the system cannot escape from it \cite{Richter-Dyn1972}.  However, there might be a regime where the proliferation of bacteria overcomes the rate of death and thus a finite stationary population density is maintained for long times. The transition between the absorbing and the active state may be continuous, with observables displaying universal scaling behaviour \cite{grassberger1997,hinrichsen2000,Odor2004,Lubeck2004,marro2005}. Although conceptually of great interest, the unambiguous observation of even the simplest non-equilibrium universality class  -- directed percolation -- is challenging and has only been achieved in recent years in a range of soft-matter systems and fluid flows \cite{Rupp2003,takeuchi2007,takeuchi2009,Lemoult2016,Sano2016,Kohl2016,takahashi2016} (see also the references in \cite{takeuchi2007,takeuchi2009}). The exploration of such universal non-equilibrium phenomena is currently an active topic across different disciplines with a number of open questions concerning, among others, their classification, the role of disorder, and quantum effects. In particular, cold atomic systems have proven to constitute a versatile platform for probing this and related physics 
\cite{Low2009,helmrich2016,Schempp:2014,Malossi:2014, Urvoy2015,Valado:2016,Simonelli:2016,letscher2016, Carr2013,Sibalic2016}.

Here we experimentally observe signatures of an absorbing state phase transition in a driven open quantum system formed by a gas of cold atoms. We laser-excite high-lying Rydberg states under so-called \emph{facilitation conditions} \cite{Ates07, Amthor2010, lesanovsky2013, lesanovsky2014}, whereby an excited atom favours the excitation of a nearby atom at a well-defined distance. This process can lead to an avalanche-like spreading of excitations \cite{Schempp:2014,Malossi:2014,Valado:2016,Simonelli:2016} and competes with spontaneous radiative decay, which drives the system towards a state without Rydberg excitations. As a result, the system displays a crossover between an absorbing state and a stationary state with a finite Rydberg excitation density. We identify signatures suggesting that this crossover is in fact a smoothed out continuous phase transition. An intriguing feature of this phase transition is that it appears to require atomic motion in order to occur in the disordered atomic gas considered here.

In our experiments we prepare cold atomic samples of ${}^{87}$Rb atoms in a magneto-optical trap (MOT) at an approximate temperature of $150\, \mu\textrm{K}$. The density distribution is Gaussian with width $\sigma = 230\ \mu\textrm{m}$ and peak density $n_0 =4.5\times10^{10}\,\textrm{cm}^{-3}$. The external driving, consisting of two co-propagating laser beams of wavelengths 420 and 1013 nm, couples the ground state $|g\rangle$ and the high-lying (Rydberg) state 70S $|r\rangle$. Atoms $i$ and $j$ in state $|r\rangle$ at positions ${\bf r}_i$ and ${\bf r}_j$ interact \cite{Amthor:2007,Comparat:10,Pfau:2012,Thaicharoen:2015,Teixeira:2015,Faoro:2016} through van der Waals interactions $V_{ij} =C_6/|{\bf r}_i-{\bf r}_j|^6$ with a positive dispersion coefficient $C_6 = h \times 869.7\ \text{GHz}\ \mu \text{m}^6$ \cite{Walker:2008}. The coupling strength between $|g\rangle$ and $|r\rangle$ is given by the (two-photon) Rabi frequency $\Omega$, and the excitation lasers can be detuned by an amount $\Delta$ from resonance.  The dephasing rate (due to the laser linewidth and residual Doppler broadening) is $\gamma = 4.4$ MHz, which is greater than the maximum value of $\Omega = 2\pi \times 250$ kHz. The system is thus in the (incoherent) strongly dissipative regime \cite{Ates2007,Petrosyan2013,Cai2013,lesanovsky2013,marcuzzi2014}. We focus on blue detuning $\Delta>0$, for which previous theoretical and experimental work \cite{lesanovsky2014,Schempp:2014,Malossi:2014,Valado:2016,Simonelli:2016} has shown that, in the presence of strong dephasing, the aforementioned facilitation mechanism increases the probability to excite (or de-excite) atoms in a spherical shell of radius $r_\text{fac} = (C_6/\hbar\Delta)^{1/6}$ around an excited atom \cite{lesanovsky2013,lesanovsky2014}.  The laser beam at 420 nm is focused to a waist of around $8\, \mu\text{m}$, which is comparable to $r_\text{fac}$ in this parameter regime, effectively reducing the excitation dynamics to one dimension (1D).

\begin{figure}[t]
\includegraphics[scale=0.5]{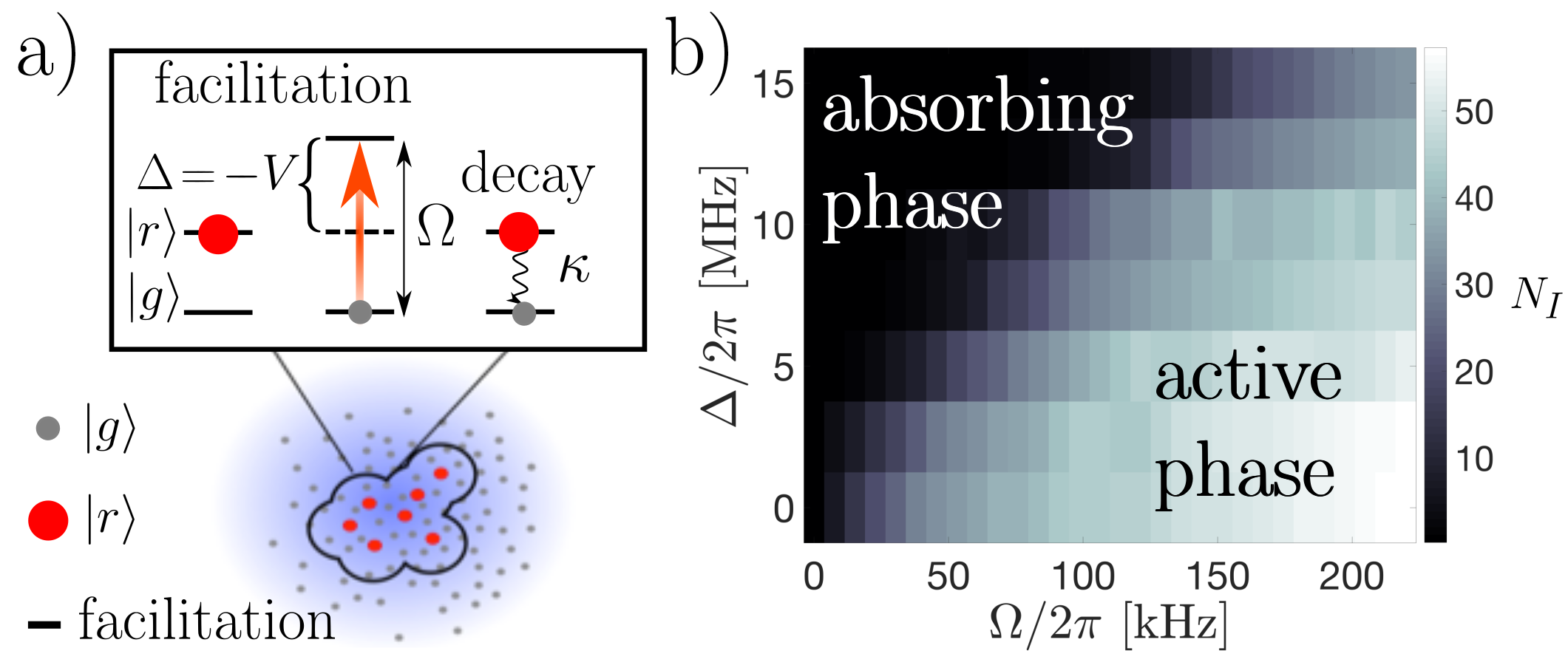}
\caption{ {\sf \bf Schematic diagram of the experimental setting and processes involved, and experimental phase diagram.}
(a) Atomic cloud with ground state atoms (gray discs), excited atoms (red discs) and atoms in the propagating facilitation region (black curved line). In the upper panel, the processes driving the dynamics are highlighted: facilitated excitations, for which the detuning $\Delta$ compensates the interaction $V$, are shown on the left ($\Omega$ is the Rabi frequency), and atomic decay on the right ($\kappa$ is the decay rate).  (b) Phase diagram showing the number of excitations $N_I$ in the stationary state as a function of $\Omega$ and $\Delta$. We observe a crossover from an absorbing state with essentially zero excitations to a fluctuating phase with a finite number of excitations.}\label{fig1}
\end{figure}

Figure \ref{fig1} (a) schematically shows the main processes occurring in our system: a configuration of ground state atoms (gray discs) and Rydberg excitations (red discs) is shown (displayed here in a 2D setting for ease of visualization), and the collective facilitation shell that results from the presence of a cluster of excitations is highlighted (black continuous line).  The dynamics is characterized by the competition between facilitation  and the spontaneous decay of excitations at a rate $\kappa = 12.5$ kHz \cite{SM}. The system eventually reaches a stationary state that depends on the relative strength of these two processes. 

Experimentally, we study the resulting stationary state by applying the following protocol. At the beginning of an experimental cycle (during which the MOT beams are switched off), 
we excite $6 \pm \sqrt{6}$ seed atoms (according to a Poissonian seed distribution) in $0.3\, \mu\text{s}$ with the excitation laser on resonance with the Rydberg transition. Thereafter, the atoms are excited at finite (two-photon) detuning $\Delta > 0$ and Rabi frequency $\Omega$ for a duration of 1.5 ms, which is much longer than the lifetime $1/\kappa$ of the 70S state \cite{SM}. Immediately after that, an elecric field is applied that field ionizes all the Rydberg atoms with principal quantum number $n \gtrsim 40$ and accelerates the ions towards a channeltron, where they are counted with a detection efficiency of 40\%. The observables of interest are based on the distribution of the number of detected ions at the end of each run. The procedure is repeated 100 times for each set of parameters, with a repetition rate of 4 Hz, in order to get reliable estimates of the mean $N_I$ and the variance $\Delta N_I {}^2$ of the number of detected ions.

In Fig. \ref{fig1} (b) we display the phase diagram resulting from this measurement procedure. The order parameter $N_I$ is plotted as a function of $\Omega$ and $\Delta$. One can clearly see a crossover between an absorbing state, with essentially zero excitations for sufficiently small $\Omega$, and a phase with a finite number of excitations for larger $\Omega$. In the remainder of this work we will focus on the nature of this crossover.

\begin{figure*}[t]
\includegraphics[scale=0.22]{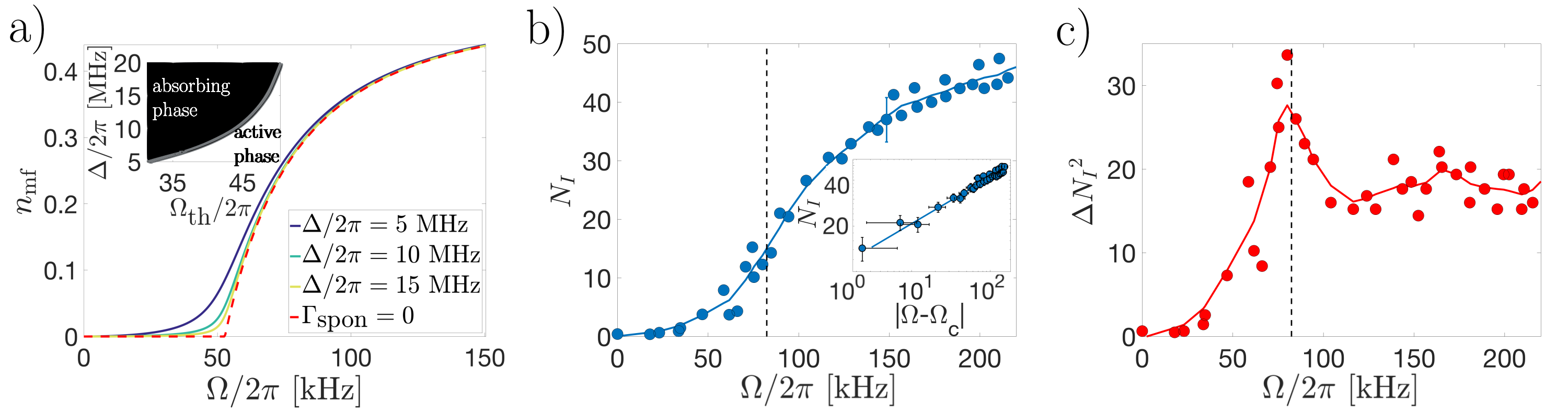}
\caption{ {\sf \bf Mean field stationary density, and experimental mean and variance of the number of excitations.} (a) Density of excitations $n_\text{mf}$ in the stationary state of the two-level mean-field model (see text) as a function of the Rabi frequency $\Omega$ for different values of the detuning $\Delta$. The correspondence between the detuning and the ratio between facilitated and spontaneous rates is as follows: for $\Delta/2\pi = 5$ MHz the ratio is $\Gamma_\text{spon}/\Gamma_\text{fac} = 19.2 \cdot 10^{-3}$,  for $\Delta/2\pi = 10$ MHz it is $\Gamma_\text{spon}/\Gamma_\text{fac} = 4.9 \cdot 10^{-3}$ and for $\Delta/2\pi= 15$ MHz it is $\Gamma_\text{spon}/\Gamma_\text{fac} = 2.2\cdot 10^{-3}$. The red dashed line shows the behavior in the absence of spontaneous (de-)excitations for $\Delta/2\pi=15$ MHz, which  shows a continuous phase transition.   The inset shows the value of $\Omega$ at which the density reaches $0.01$ -- which we denote $\Omega_\text{th}$ -- as a function of $\Delta$. (b) Average number of excitations at the end of the $1.5$ ms time window in the experiment for $\Delta/2\pi= 10$ MHz. One representative error bar is shown, corresponding to one standard deviation. {\it Inset}: same data in loglog plot for $\Omega > \Omega_c = 2\pi \times  (82.4 \pm 0.2)$ kHz. A power law nonlinear fit based on the expression $\log(N_I)=\alpha + \beta \log(\Omega - \Omega_c)$ has been applied to the data, yielding an exponent $\beta = 0.31 \pm 0.04$. The horizontal error bars correspond to a relative uncertainty of $\pm 5\%$ in the measurement of $\Omega$ due to fluctuations in the laser intensity, and possible misalignments of the beams.  The vertical error bars correspond to the measured standard deviations of the number of excitations. (c) Variance of the number of excitations as a function of $\Omega/2\pi$ based on the same experimental data. The continuous line in panels (b) and (c) is a guide to the eye and results from a sliding average, and the dashed vertical lines indicate the position of the critical point.}\label{fig2}
\end{figure*}

To provide some qualitative theoretical insight, we first conduct a simple mean-field analysis based on a 1D system that follows the same dynamical rules. Adopting the semi-classical description of Ref.~\cite{lesanovsky2014}, the (de-)excitation of atom $i$ occurs at a rate $\Gamma_i$ that depends on the configuration of neighbouring excitations.  If we neglect the correlations between atoms, the average $\langle n_i\rangle$ of the number operator $n_i \equiv |r\rangle_i\langle r|$ acting on site $i$ evolves in time according to
\begin{eqnarray}
\partial_t \langle n_i(t) \rangle  &=&  \langle - | \Gamma_i (1-2 n_i) |P(t)\rangle - \kappa \langle n_i(t) \rangle,
\label{deposition_eq}
\end{eqnarray}
where $|P(t)\rangle \equiv \sum_\mathcal{C} P(\mathcal{C};t) |\mathcal{C}\rangle$, the kets $|\mathcal{C}\rangle$ are the classical atomic configurations in the number basis (the eigenbasis of all the $n_i$), $P(\mathcal{C};t)$ is the probability of configuration $|\mathcal{C}\rangle$ at time $t$, and $|-\rangle \equiv  \sum_\mathcal{C} |\mathcal{C}\rangle$. At this point we introduce the simplifying assumption that the rate $\Gamma_i$ can take only two values: the facilitated rate $\Gamma_\text{fac}$ 
if the i-th atom lies in the facilitation shell of an existing excitation, or otherwise the spontaneous rate $\Gamma_\text{spon}$,
corresponding to the rate in the absence of nearby excitations,
\begin{equation}
\Gamma_\text{fac} \equiv \Omega^2/2\gamma;\ \Gamma_\text{spon} \equiv \left(\Omega^2/2\gamma\right)\left[1+\Delta^2/\gamma^2\right]^{-1}.
\end{equation}
In a coarse-grained description of the system, where $n  \equiv N_\mathcal{V}^{-1} \sum_{i\in\mathcal{V}} n_i$ is the fraction of excited atoms in a region of space $\mathcal{V}$ (spanning a few facilitation radii) with $N_\mathcal{V}$ atoms in it, we expect the average rate to be $n\, \Gamma_\text{fac} + (1-n)\, \Gamma_\text{spon}$. Assuming homogeneity, the spatially averaged dynamics is given by
\begin{equation}
\dot{n} = \Gamma_\text{fac} n (1 - 2 n) +  \Gamma_\text{spon} (1-n) (1 - 2 n) -\kappa n.
\end{equation}
We first consider the limit $\Gamma_\text{spon}/\Gamma_\text{fac} \to 0$ (i.e. $\Delta/\gamma \to \infty$), where the dynamics is purely governed by the competition between facilitation and decay. The stationary state solution for $\Gamma_\text{fac}<\kappa$ is the state without excitations, which constitutes an absorbing state of the dynamics. For $\Gamma_\text{fac}\geq\kappa$ facilitation prevails over decay, and the absorbing state becomes unstable, leading to a finite density stationary state,
\begin{equation}
n_\text{mf} = 
     \begin{cases}
       0, &\quad\text{if}\ \Gamma_\text{fac}<\kappa, \\
       (1- \kappa/\Gamma_\text{fac})/2, &\quad\text{otherwise.} \\
     \end{cases}
\end{equation}
As $n_\text{mf}$ is continuous at $\Gamma_\text{fac}=\kappa$, but its first derivative with respect to $\Gamma_\text{fac}$ is not, this indicates the existence (at the mean-field level) of a non-equilibrium continuous phase transition between an absorbing state with zero excitations and a fluctuating phase with a finite density \cite{hinrichsen2000}. 
Since in our experiment atoms in the 70S state can migrate (via black-body radiation) to other Rydberg states, we additionally devised a three-level model taking into account this effect, which shows the same qualitative behavior (see \cite{SM}).


In Fig. \ref{fig2} (a) we plot $n_\text{mf}$ as a function of the Rabi frequency $\Omega$ for different values of the detuning $\Delta$, using the experimental values of the dephasing and decay rates.  For the largest value of $\Delta$, we also explore the stationary state in the absence of spontaneous excitations, $\Gamma_\text{spon}=0$  (see the red dashed line), which shows the aforementioned phase transition. For non-vanishing $\Gamma_\text{spon}/\Gamma_\text{fac}$, $n_\text{mf}$ is always positive and the non-analyticity at $\Gamma_\text{fac} = \kappa$ is smoothed out into a crossover (see the continuous lines). For larger values of $\Delta$, as $\Gamma_\text{spon}$ is suppressed, the system is expected to be closer to the critical point. In the inset, we show the position of the threshold $\Omega_\text{th}$, which we set to be the value of the Rabi frequency for which $n_\text{mf} = 0.01$. We take this to be an approximate measure of the onset of the crossover between the absorbing phase and the active phase away from the critical point. 
We conjecture that the same physics lies at the basis of the phase diagram in Fig.~\ref{fig1}, which would thus signal the presence of a smoothed phase transition in the experiment. The smoothness stems from the spontaneous rate $\Gamma_\text{spon}$ which acts
like a field that off-sets the system away from criticality. By substituting our estimates of the experimental parameters, we find $\Gamma_\text{spon}/\Gamma_\text{fac}$ to be of the order of $10^{-3}$ for $\Omega/2\pi =125\ \mathrm{kHz}$ and $|\Delta/2\pi|=10\ \mathrm{MHz}$ \cite{SM}.

In the presence of a continuous phase transition, we would expect the experimental data to show a smoothed-out singularity in the fluctuations and a power-law behavior in the number of excitations \cite{hinrichsen2000}. This is, indeed, compatible with what we observe.  
In Fig.~\ref{fig2} (b) the number of excitations $N_I$ is plotted as a function of $\Omega$ for a fixed detuning $\Delta = 2\pi \times 10$ MHz. The continuous line results from a sliding average, and is meant as a guide to the eye. In  Fig. \ref{fig2} (c) we show the variance of the number of excitations $\Delta N_I {}^2$ for the same data as in (b), which displays a clear peak around $\Omega/2\pi = 80$ kHz. Approaching a critical point, the correlation length diverges, and global density fluctuations should correspondingly diverge.  In the inset of Fig. \ref{fig2} (b), $N_I$ is plotted on a reduced interval in logarithmic scale. Since the position of the peak gives the approximate location of the critical Rabi frequency, $\Omega_c$ is chosen in its neighborhood as the value that maximizes the goodness of the nonlinear fit. This procedure yields a value of $\Omega_c = 2 \pi \times (82.4\pm 0.2)$ kHz [dashed vertical line in Fig. \ref{fig2} (b) and (c)] and a power-law dependence $N_I \sim (\Omega-\Omega_c)^\beta$ with an exponent $\beta \approx 0.31 \pm 0.04$ (see below a discussion of the significance of this result).

We turn now to a closer inspection of the role of disorder in the atomic cloud. This will highlight the relevance of atomic motion as a central ingredient for the observed physics \cite{Sibalic2016}. In order to undergo a phase transition, the system must establish correlations over mesoscopic length scales, and to analyze whether this is possible we have to consider two experimental features that so far have not been discussed: positional disorder and atomic motion.  To this end, we use an effective 1D lattice model comprising $L$ sites occupied by $N$ atoms ($L>N$) located at random positions. We first address the hypothetical situation in which the positions are frozen for the duration of the experiment, so that the spatial configuration induces {\it quenched} disorder on the excitation rates. A prerequisite for the formation of a large cluster of excitations is the existence of a large number of atoms located at a distance $r_\text{fac}$ from each other, and a simple argument shows that the probability of finding such regularly-spaced clusters is exponentially suppressed in their size \cite{SM}. For example, if we estimate the effective length of the cloud to be the distance between the positions at which the density drops to $1\%$ of the value at the peak (on either side), which gives $L_\textrm{eff} \simeq 990\ \mu\textrm{m}$, and if we consider there are $k=10$ sites per $r_\text{fac}$, the experimental conditions translate into a density $\rho \equiv N/L \approx 0.3$. Under these conditions, the resulting probability of occurrence of an occupied sublattice of size $L_\text{eff}/10 \approx 15\, r_\text{fac}$ is considerably smaller than $10^{-6}$. This illustrates the fact that correlations over mesoscopic length scales are extremely unlikely to develop in the cloud.

\begin{figure}[t]
\hspace{-0.5cm}\includegraphics[scale=0.13]{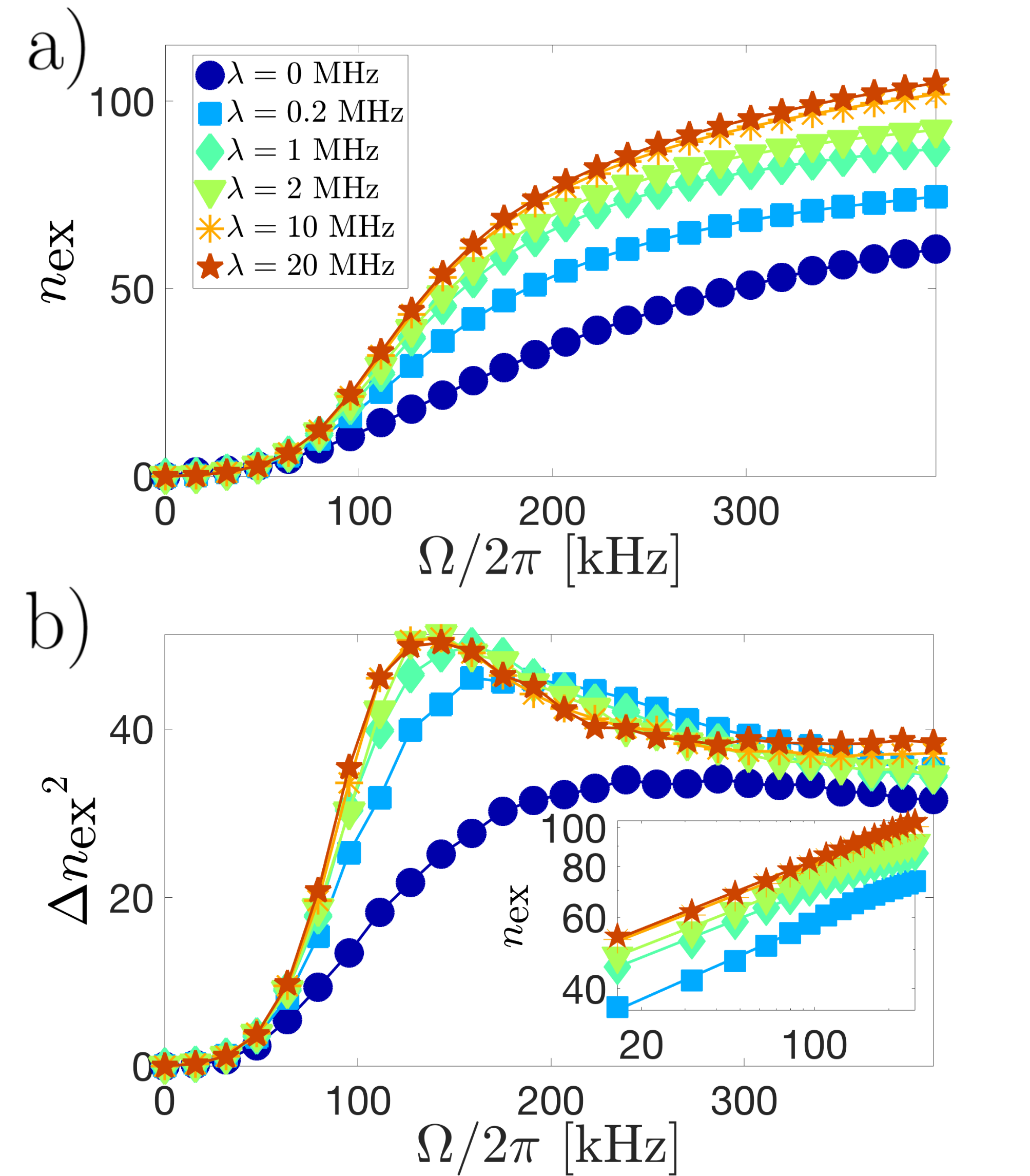}
\caption{ {\sf \bf Mean and variance of the number of excitations as a function of $\Omega/2\pi$ in a 1D model with atomic motion.} Results based on a chain of $L=1500$ sites and $N=450$ atoms with the experimental laser and atomic level parameters, and a range of mobility $\lambda$ based on the experimental atomic motion.
(a) Mean number of excitations $n_\text{ex}$ as a function of $\Omega/2\pi$ for mobilities $\lambda = 0$ (quenched disorder), $0.2$, $1$, $2$, $10$, $20$ MHz. (b) Fluctuations of the excitation number $\Delta n_\text{ex} {}^2$ as a function of $\Omega/2\pi$ for different $\lambda$ [color code and markers as in panel (a)]. The inset shows a logarithmic plot of $n_\text{ex}$ vs $(\Omega-\Omega_c)/2\pi$, where $\Omega_c$ is defined to be the value of $\Omega$ where the fluctations reach a peak, and associated power law fits.\label{fig3}}
\end{figure}

However, in our experiment the timescales are too long for this frozen gas picture to hold. In fact, the mean atomic velocity of our samples translates into a mean displacement of around 0.19 m/s (for $T = 150\ \mu\text{K}$), meaning that on the timescale of an experimental cycle an atom can traverse a distance comparable to the width of the cloud. The excitation dynamics proceeds on an ever changing background, which corresponds to {\it annealed} disorder. To study this effect we use the lattice model discussed above, with the atomic motion parametrized by the mobility $\lambda$, which is the rate at which atoms jump to neighboring sites (so long as they do not violate the single occupancy condition). The inclusion of these jump processes is a minimal way to account for thermal motion as well as mechanical effects due to repulsion between Rydberg states. For the spreading of excitations to become possible, atomic motion should act in such a way that excitations have at least an atom going through their facilitation shell before decaying.

In Fig.~\ref{fig3} (a) and (b) we plot the mean number of excitations $n_\text{ex}$ and the variance $\Delta n_\text{ex} {}^2$, respectively, as a function of $\Omega$ for a chain with $k=10$ sites per facilitation distance $r_\textrm{fac}$ and $L = k L_\text{eff}/r_\text{fac}  = 1500$. The density of occupied sites of choice, $\rho = 0.3$ ($N=450$), and the range of $\lambda$ values considered [see panel (a) for the color coding] have been adjusted to match the experimental conditions (see \cite{SM}), while the rest of the parameters are those of the experiment (with $\Delta = 2\pi \times 10\,  \textrm{MHz}$). For $\lambda = 0$ (quenched disorder) the growth of $n_\text{ex}$ with $\Omega$ is mild and the fluctuations $\Delta n_\text{ex} {}^2$ do not display a peak. As $\lambda$ is increased (i.e., for time-dependent disorder), however, the growth becomes more abrupt and the fluctuations display a clear peak.   In the inset of  Fig.~\ref{fig3} (b) we include a logarithmic plot of $n_\text{ex}$ against $(\Omega-\Omega_c)$ for $\lambda>0$, where $\Omega_c$ is the position of the peak. The results are compatible with a power law dependence $n_\text{ex} \sim (\Omega-\Omega_c)^\beta$, especially for large mobilities, with an exponent that appears to saturate around $\beta \approx 0.25 \pm 0.04$. From this we conclude that in our model, and probably in our experimental system, atomic motion proves crucial for the emergence of pronounced fluctuations and scaling behavior. 

In summary, we have presented experimental data that show a crossover between an absorbing phase without Rydberg excitations and an active phase with a finite fraction of Rydberg excitations in an open dissipative atomic gas. Evidence for the existence of a underlying non-equilibrium continuous phase transition has been provided. In fact, the  effective mean-field model as well as the extracted scaling exponent suggest a connection to directed percolation (DP), which is one of the simplest non-equilibrium universality classes. DP has previously been predicted to emerge in Rydberg lattice systems \cite{marcuzzi2015}. The scaling exponent extracted from the experimental data is compatible with that of DP in one dimension, $\beta_\text{DP} = 0.276486(8)$ \cite{hinrichsen2000}. A crucial issue of the current experiment is the nature and role of disorder. The point we have emphasized above, namely that quenched disorder heavily distorts the critical behavior, whereas annealed disorder does not, has been established for DP via field-theoretical and numerical approaches  \cite{Janssen1997,Cafiero1998,Moreira1996,hinrichsen2000}. 
A future goal is to fully characterize and classify the non-equilibrium phases of driven Rydberg gases, e.g., through more precise measurements of static and dynamic exponents and also a field-theoretical study of the universal properties. An exciting perspective is that Rydberg gases allow the controlled inclusion of quantum effects, e.g., by reducing the dephasing rate. Future studies will thus potentially access new dynamical regimes that go beyond the current body of knowledge on out-of-equilibrium phase transitions, which is largely focused on classical many-body systems \cite{hinrichsen2000,Odor2004}.

\begin{acknowledgements}
\emph{Acknowledgements ---} RG, MM and IL would like to thank Juan P. Garrahan and Carlos P\'erez-Espigares for useful discussions. The research leading to these results has received funding from the European Research Council under the European Union's Seventh Framework Programme (FP/2007-2013) / ERC Grant Agreement No. 335266 (ESCQUMA), the EU-FET grant HAIRS 612862 and from the University of Nottingham. Further funding was received through the H2020-FETPROACT-2014 grant No.  640378 (RYSQ). RG acknowledges the funding received from the European Union's Horizon 2020 research and innovation programme under the Marie Sklodowska-Curie grant agreement No. 703683. We also acknowledge financial support from EPSRC Grant no.\ EP/M014266/1. Our work has benefited from the computational resources and assistance  provided  by the University of Nottingham High Performance Computing service.
\end{acknowledgements}


%

\onecolumngrid
\newpage

\renewcommand\thesection{S\arabic{section}}
\renewcommand\theequation{S\arabic{equation}}
\renewcommand\thefigure{S\arabic{figure}}
\setcounter{equation}{0}
\setcounter{figure}{0}

\begin{center}
{\Large Supplementary Material: Experimental signatures of an absorbing-state phase transition in an open driven many-body quantum system}
\end{center}

\section{Time-dependent data and experimental decay rates}

\begin{figure}[h!]
\includegraphics[scale=.3]{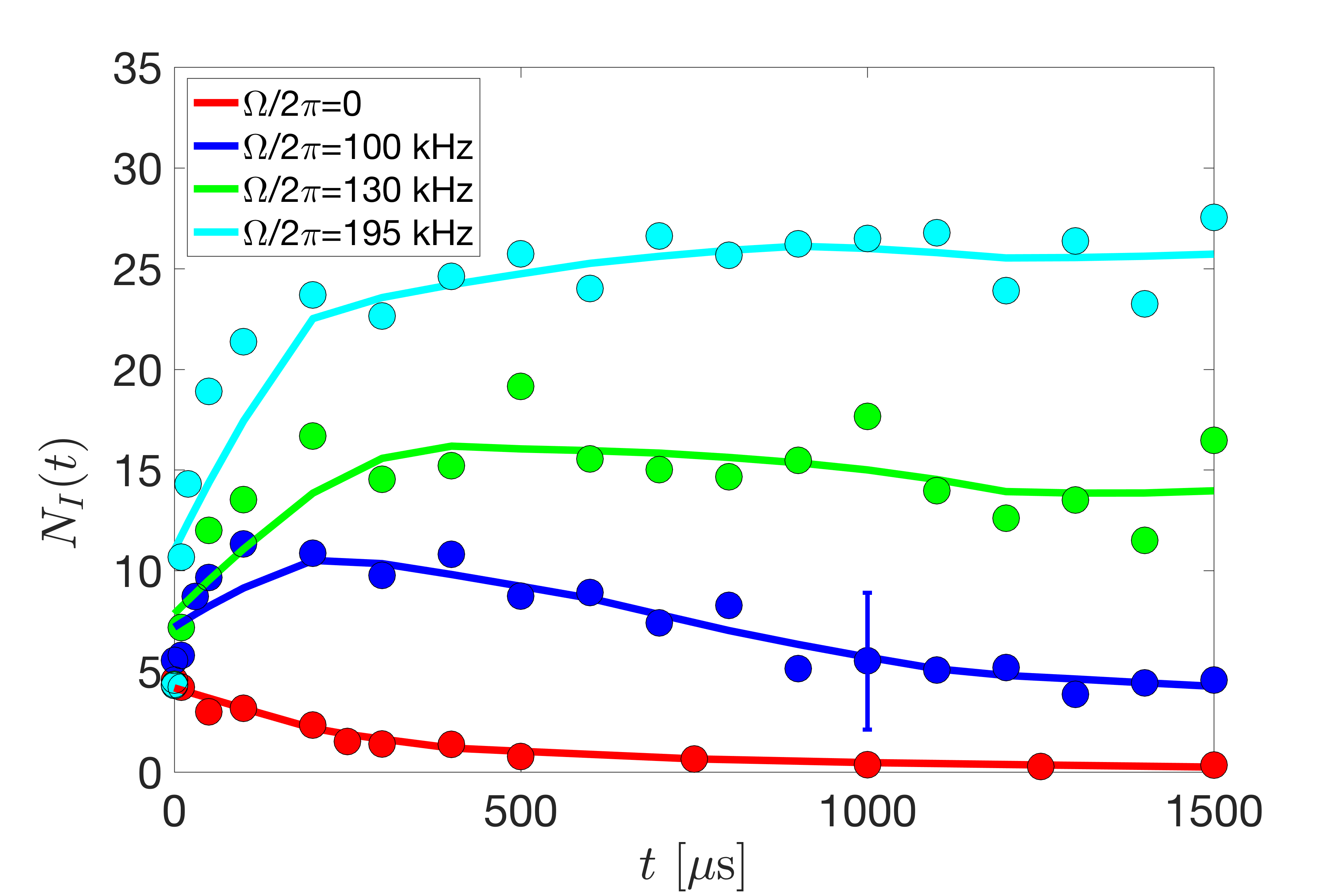}
\caption{ {\sf \bf Time-dependent data for different Rabi frequencies $\Omega$ and detuning $\Delta/2\pi = 10$ MHz.}
Number of detected ions $N_I(t)$ during the 1.5-ms-excitation time window. Continuous lines based on a sliding average. Different colors correspond to different values of $\Omega$ (see legend). A representative error bar is shown, corresponding to one standard deviation.}\label{fig0SM}
\end{figure}

In the main text we show data that result from exciting the atomic gas at a Rabi frequency $\Omega$ ranging from 0 to $2\pi \times 250$ kHz for various detuning values $\Delta$ during a time window of 1.5 ms.  In Figure \ref{fig0SM} we show typical curves of the mean number of detected ions $N_I(t)$ (which is proportional to the number of Rydberg excitations) as a function of the excitation time for $\Delta/2\pi=10$ MHz. To obtain these data, first we excite $6 \pm \sqrt{6}$ seed atoms (according to a Poissonian seed distribution) during a time window of $0.3\, \mu\text{s}$, then apply the out-of-resonance laser field to the MOT comprising 1 660 000 atoms with peak density $4\cdot 10^{10}\, \text{cm}^{-3}$. Discs of different colors correspond to different values of $\Omega$, and the continuous lines are meant as a guide to the eye. The time window appears to be long enough to consider that, by the end of it, the system is at -- or at least close to -- the stationary state. The same conclusion can be drawn from the time-dependent data (not shown) corresponding to the other detuning values included in the phase diagram of Fig. 1 (b) in the main text.

The 1.5-ms experimental time window is much larger than the experimentally determined lifetime of the 70S state, which is measured by using a depumping technique \cite{Simonelli2016DepSM} that selectively depumps the 70S state via the fast decaying 6P state. This allows us to determine, as a function of time after the excitation pulse, both the total Rydberg population with $n\gtrsim 40$ by field ionzation as well as the population of the 70S state alone, for which we find a lifetime $\tau \approx 80\, \mu\text{s}$. For the purposes of this work we assume that the migration from the 70S state to nearby Rydberg states by black-body radiation essentialy works as a loss mechanism from the 70S state and should not influence the main features of the crossover behavior we observe, especially close to the critical point, where the number of excitations is small.

\section{Effective dynamics of a dissipative Rydberg gas}

Previous theoretical work \cite{lesanovsky2013SM,marcuzzi2014SM,lesanovsky2014SM} has shown that in a driven system of Rydberg atoms the presence of strong dephasing noise leads to a rapid decay of the off-diagonal elements of the density matrix, which results in an effective dynamics that proceeds along classically accessible states and is governed by a classical master equation. The starting point is an open quantum description with a spin Hamiltonian 
\begin{equation}
	H =  \sum_k \left[  \frac{\Omega}{2} \sigma_x^k + \Delta n_k + \frac{1}{2} \sum_{q\neq k} \frac{C_6}{ |r_q - r_k|^6} n_k n_q    \right] ,
\end{equation}
where $\Omega$ is the Rabi frequency, $\gamma$ is the dephasing rate, $\Delta$ is the value of the detuning and $C_6$ is the dispersion coefficient of the van der Waals interactions, and Lindblad jump operators $L_{1,k} = \sqrt{2\gamma} n_k$ for dephasing and $L_{2,k} = \sqrt{\kappa} \sigma_-^k$ for decay, with $2 \gamma$ and $\kappa$ the corresponding rates. The quantum operator $n_k$ is shorthand for $(\sigma_z^k + 1)/2$ and corresponds to the projector onto the $\left|{\uparrow}_k \right\rangle$ state. For $\gamma \gg \Omega$, one can derive the effective rate equation
\begin{equation}
\partial_t |P(t)\rangle = \sum_{k=1}^N \Gamma_k \left[\sigma_+^k - (1 - n_k) \right] |P(t)\rangle 
\label{dynamics} +\sum_{k=1}^N (\Gamma_k+\kappa) \left[\sigma_-^k - n_k \right] |P(t)\rangle 
\end{equation} 
where $|P(t)\rangle \equiv \sum_\mathcal{C} P(\mathcal{C};t) |\mathcal{C}\rangle$, $P(\mathcal{C};t)$ being the probability of configuration $|\mathcal{C}\rangle$ at time $t$. Here,  $|\mathcal{C}\rangle$ can be any classical configuration such as $|\!\uparrow \downarrow \downarrow \uparrow \cdots \uparrow\rangle$.  The operator $n_k$ acting on $|\mathcal{C}\rangle$  amounts to a multiplication by one if atom $k$ is excited and by zero if it is not. The (de)excitation operator $\sigma_+^k$ ($\sigma_-^k$) acting on $|\mathcal{C}\rangle$ creates (annihiliates) an excitation if $k$ was in the ground state (Rydberg state) in $\mathcal{C}$, yielding zero otherwise.  While the decay rate $\kappa$ is constant, the configuration-dependent operator-valued (de)excitation rates are given by
\begin{equation}
\Gamma_k=  \frac{\Omega^2}{2\gamma}\left[1+\left(\frac{\Delta -  C_6 \sum_{q\neq k} \frac{n_q}{|r_q - r_k|^6}}{\gamma} \right)^2\right]^{-1}.
\label{rates}
\end{equation} 

\begin{figure}[h!]
\includegraphics[scale=0.15]{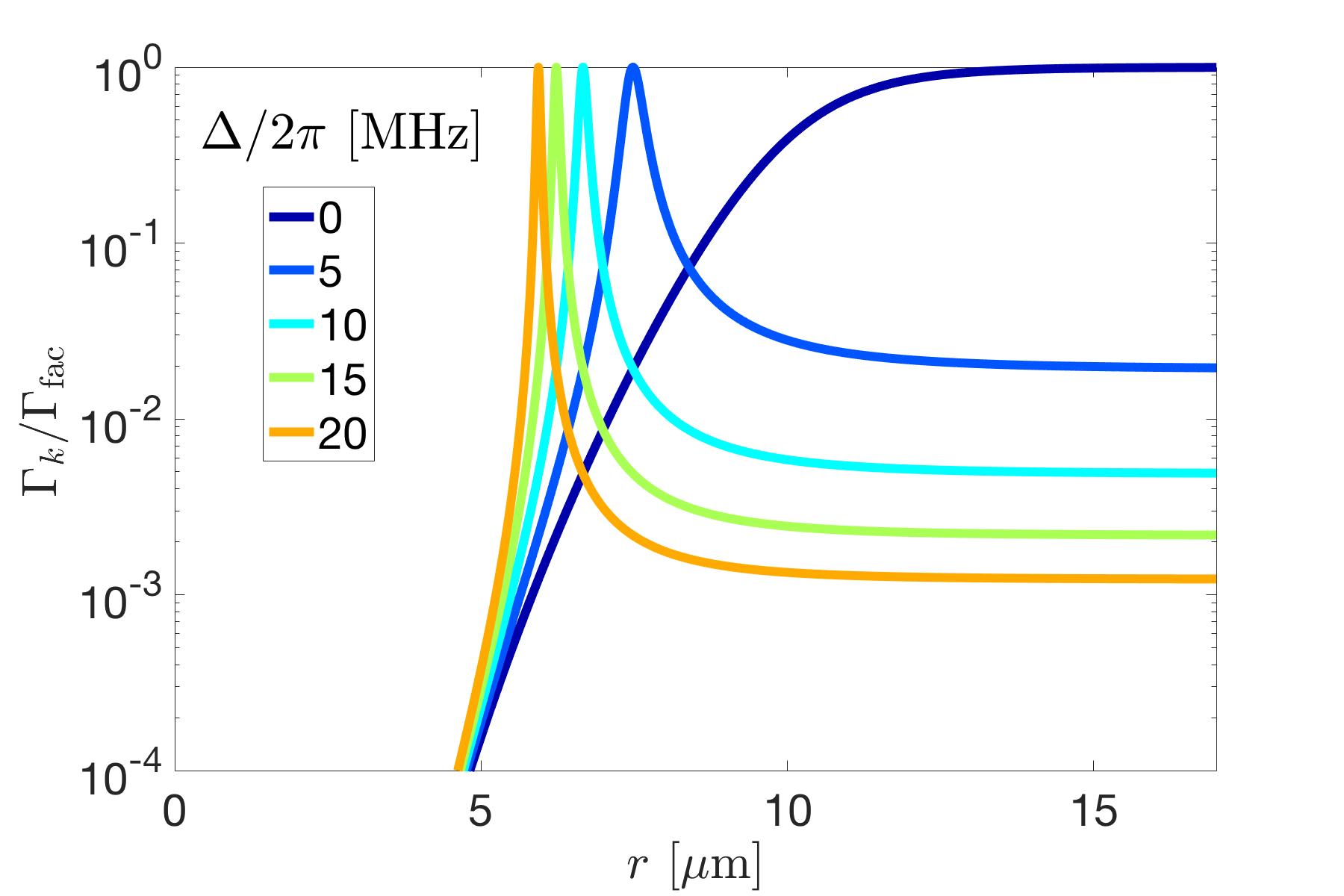}
\caption{ {\sf \bf Excitation rates for different values of the detuning.}
$\Gamma_k/\Gamma_\text{fac}$ as a function of the distance to a Rydberg excitations $r$ for different detuning values $\Delta$ (see legend).}\label{fig1SM}
\end{figure}

The detuning $\Delta$ plays a vital role in the excitation dynamics. This is most clearly seen in a two-atom configuration, by  inspecting the excitation rate of an atom as a function of its distance $r$ to an excited atom in the Rydberg state.  When the distance is equal to the {\it facilitating radius} $r = r_\text{fac} \equiv \left(C_6/\hbar\Delta\right)^{1/6}$, the rate reaches a maximum $\Gamma_\text{fac} \equiv \Omega^2/2\gamma$. In Fig. \ref{fig1SM} we show $\Gamma_k$ normalized by $\Gamma_\text{fac}$ as a function of $r$ for the experimental values $C_6 = h \times 869.7 \ \text{GHz}\ \mu \text{m}^6$ and $\gamma = 2\pi \times 700$ kHz, and several values of $\Delta$. It is clear how the larger the detuning, the narrower the facilitating region and the smaller the rate for a spontaneous (de)excitation $\Gamma_\text{spon} \equiv \frac{\Omega^2}{2\gamma}/\left[1+\frac{\Delta^2}{\gamma^2}\right]$ that is achieved for large $r$.

We experimentally determine $\Gamma_\text{spon}$ by off-resonantly exciting Rydberg atoms with negative rather than positive detuning. In this way, there is no facilitation mechanism (for the repulsive 70S state used in our experiments), and hence only single excitations are created at a rate $\Gamma_\text{spon}$. From the total rate of excitations measured we obtain the single-particle value by dividing by the number of atoms in the interaction volume defined by the size of the MOT and the waist of the blue leaser beam. In order to determine $\Gamma_\text{spon}$ close to the critical point for the data reported in the main text, we performed this experiment with $\Omega/2\pi= 125$ kHz and $\Delta/2\pi = -10$ MHz and
found $\Gamma_\text{spon} = 2\pi\times 5$ Hz. This value is smaller than the theoretically predicted value of approx. $2\pi \times 27$ Hz, probably due to systematic errors in the measurements of $\Omega$ and the number of atoms as well as imperfect alignment of the laser beams. To compare with the rate of facilitated events, we assume an intermediate value of $2\pi \times 10$ Hz, which yields $\Gamma_\text{spon}/\Gamma_\text{fac} = 9\cdot 10^{-4}$. 

\section{Mean-field behavior in the presence of an additional Rydberg level}
 
\begin{figure}[h!]
  \centering
    \includegraphics[width=0.7\textwidth]{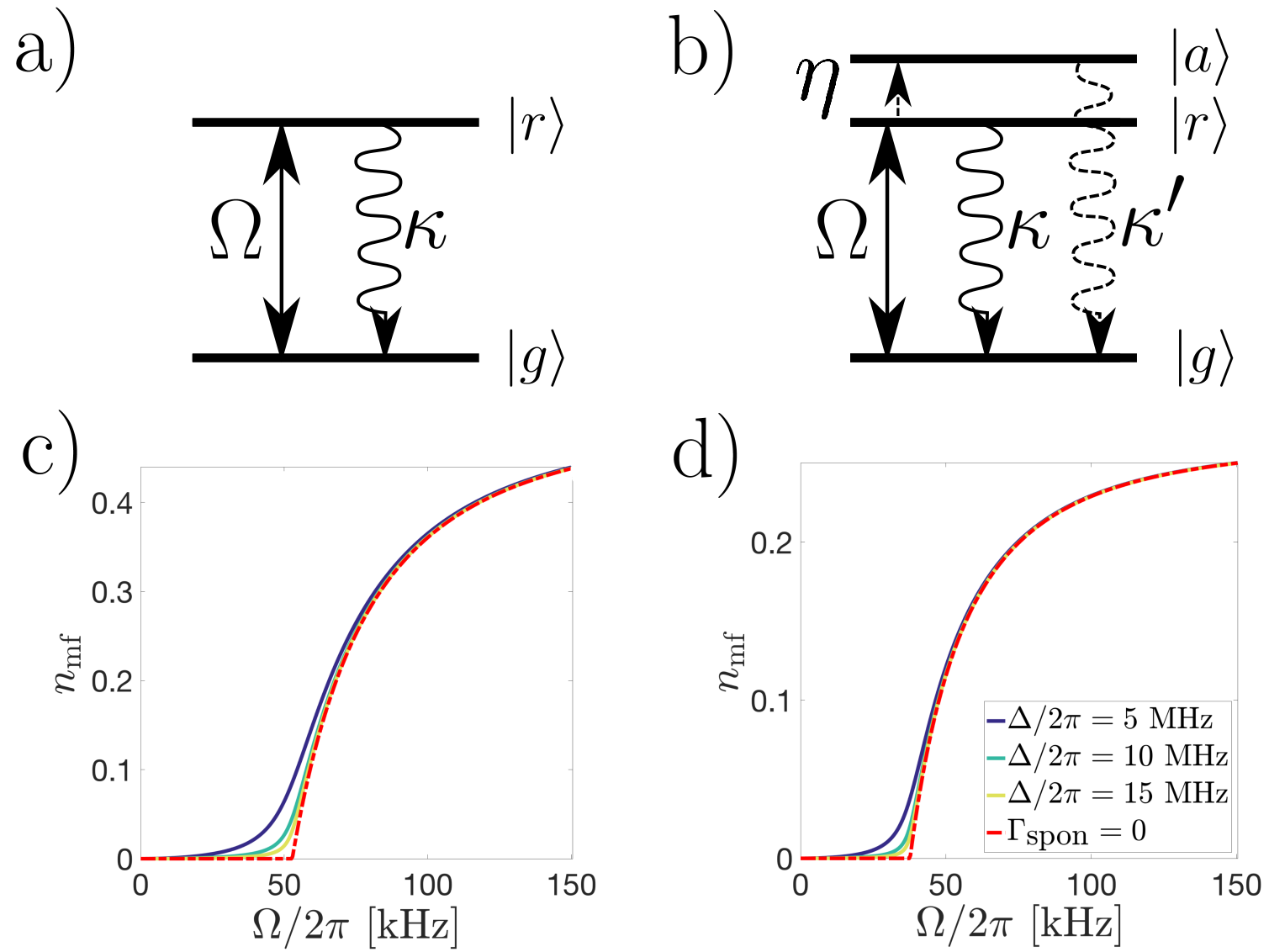}
\caption{{\bf Level schemes, and mean-field  stationary density of excitations in the two-level situation considered in the main text and when including transitions to a nearby Rydberg level.} (a) Two-level system comprising a ground state $|g\rangle$ and a Rydberg state $|r\rangle$. $\Omega$ is the Rabi frequency and $\kappa$ the decay rate. (b) Three-level system where transitions to an additional Rydberg level $|a\rangle$ are considered. The transition rate from $|r\rangle$ to $|a\rangle$ is given by $\eta$, while $\kappa^\prime$ is the decay rate of the additional state. (c) Density of excitations $n_\text{mf}$ in the stationary state of the mean-field two-level model (see text) $n_\text{mf}$ as a function of the Rabi frequency $\Omega$ for different values of the detuning $\Delta$ [see legend in panel (d)]. These data are included in Fig. 2 (a) in the main text. The correspondence between the detuning and the ratio between facilitated and spontaneous rates is as follows: for $\Delta/2\pi = 5$ MHz the ratio is $\Gamma_\text{spon}/\Gamma_\text{fac} = 19.2 \cdot 10^{-3}$,  for $\Delta/2\pi = 10$ MHz it is $\Gamma_\text{spon}/\Gamma_\text{fac} = 4.9 \cdot 10^{-3}$ and for $\Delta/2\pi= 15$ MHz it is $\Gamma_\text{spon}/\Gamma_\text{fac} = 2.2\cdot 10^{-3}$. The red dashed line shows the behavior in the absence of spontaneous (de-)excitations for $\Delta/2\pi=15$ MHz, which shows a continuous phase transition. (d) Density of excitations $n_\text{mf}$ in the stationary state of the mean-field three-level model. The parameters of choice for processes involving $|a\rangle$ are the transition rate $\eta = 5$ kHz and the decay rate $\kappa^\prime = 2.86$ kHz (see text for an explanation), and the rest are the same as those used in panel (c). The vertical axis has a different scale from that used in panel (c). The detected population of excited atoms in the experiment is $n_\text{mf} + m_\text{mf} = (1 + \eta/\kappa^\prime)\, n_\text{mf}$, which is $2.75\,  n_\text{mf}$ for the parameters of choice. }
\label{fig3levels}
\end{figure}

In the mean-field treatment included in the main text, where only the ground state $|g\rangle$ and a Rydberg state $|r\rangle$ are considered, see Fig. \ref{fig3levels} (a), the time evolution of the density of excited atoms $n$ follows 
\begin{equation}
\dot{n} = \Gamma_\text{fac} (1-n) n  - \Gamma_\text{fac} n^2+ \Gamma_\text{spon} (1- n)^2 - \Gamma_\text{spon} n (1-n)  - \kappa n,
\label{eq2levels}
\end{equation}
where $\Gamma_\text{fac}$ is the (de-)excitation rate of facilitated atoms, $\Gamma_\text{spon}$ is the (de-)excitation rate in the absence of facilitating excitations ($\Gamma_\text{fac} > \Gamma_\text{spon}$), and $\kappa$ is the decay rate. A detailed explanation is provided in the main text, where this equation appears (after a slight rearrangement) as Eq. (2). In the stationary state, the density of excitations is
\begin{equation}
n_\text{mf} = \frac{\Gamma_\text{fac} - 3 \Gamma_\text{spon} - \kappa + \sqrt{\Gamma_\text{fac}^2+ 2 \Gamma_\text{fac} \Gamma_\text{spon} +\Gamma_\text{spon}^2 - 2 \Gamma_\text{fac} \kappa + 6 \Gamma_\text{spon} \kappa + \kappa^2}}{4(\Gamma_\text{fac}-\Gamma_\text{spon})},
\label{eq2levelfixedp}
\end{equation}
\noindent which is always positive and is a stable fixed point of Eq. (\ref{eq2levels}). In the absence of spontaneous excitations ($\Gamma_\text{spon} = 0$), the stable stationary solution adopts the following form:
\begin{equation}
n_\text{mf} = \left. \begin{cases} 
      \displaystyle 0 & \Gamma_\text{fac}\leq \kappa \\
      \displaystyle\frac{\Gamma_\text{fac}-\kappa}{2 \Gamma_\text{fac}} &  \Gamma_\text{fac} > \kappa.  \\ 
   \end{cases}
\right.\label{eq2levelsfixedp_withoutspon}
\end{equation}

\noindent While Eq. (\ref{eq2levelsfixedp_withoutspon}) reflects a continuous phase transition, Eq. (\ref{eq2levelfixedp}) shows the crossover that is observed in the presence of spontaneous excitations ($\Gamma_\text{spon} > 0$), see Fig. \ref{fig3levels} (a) (which displays the main results shown in Fig. 2 (a) in the main text, and is included here again for convenience).

In our experiment, due to the presence of blackbody radiation excited atoms can make a transition to nearby excited levels. Here we try to account for this effect in our mean-field description and show that this does not change the fundamental physics.
We thus introduce an auxiliary Rydberg level $|a\rangle$, which effectively accounts for all the nearby levels in which atoms can end up due to this effect. This leads to two additional processes: transitions from $|r\rangle$ to $|a\rangle$ can occur with a given rate $\eta$, and atoms in state $|a\rangle$ can decay to the ground state with a rate $\kappa^\prime$, see Fig. \ref{fig3levels} (b). Furthermore, as level $|a\rangle$ is expected to lie close to $|r\rangle$, we assume that atoms excited to $|a\rangle$ also have the ability to facilitate the $|g\rangle \leftrightarrow |r\rangle$ transition for nearby atoms.  The equations of motion for the numbers $n$ of atoms in $|r\rangle$, and $m$ of atoms in $|a\rangle$ are respectively
\begin{align}
\label{eq3levels1}
     \dot{n} & = \Gamma_\text{fac} (1-(n+m))(n+m) - \Gamma_\text{fac} n (n+m) + \Gamma_\text{spon} (1-(n+m))^2 - \Gamma_\text{spon} n (1-(m+n)) - (\kappa +\eta) n  \\
\label{eq3levels2} \dot{m} & = \eta\, n - \kappa^\prime m, 
\end{align}
\noindent while the time derivative of the number of atoms in the ground state is $-(\dot{n} + \dot{m})$ by probability conservation ($n+m+g=1$). In the stationary state $m_\text{mf} = \frac{\eta}{\kappa^\prime} n_\text{mf}$. By plugging this expression into Eq. (\ref{eq3levels1}), we find the following stationary value for $n$
\begin{equation}
n_\text{mf}\! =\! \frac{\Gamma_\text{fac}\! \left(1\! +\! \frac{\eta}{\kappa^\prime}\right)\! -\! \Gamma_\text{spon}\! \left(3\!+\!\frac{2\eta}{\kappa^\prime}\right)\! -\! \kappa\! -\! \eta\!
 +\! \sqrt{\left(\Gamma_\text{fac}\! \left(\!1\! +\! \frac{\eta}{\kappa^\prime}\right)\! -\! \Gamma_\text{spon} \!\left(3\!+\!\frac{2\eta}{\kappa^\prime}\right)\! -\! \kappa\! -\! \eta\right)^2\! +\! 4 \Gamma_\text{spon}\left(\Gamma_\text{fac}\! -\! \Gamma_\text{spon}\right) \left(2\! +\! \frac{3\eta}{\kappa^\prime}\! +\! \frac{\eta^2}{\kappa^{\prime2}}\right) }}{2\left(\Gamma_\text{fac} - \Gamma_\text{spon}\right) \left(2 + \frac{3\eta}{\kappa^\prime} + \frac{\eta^2}{\kappa^{\prime2}}\right)},
\label{eq3levelfixedp}
\end{equation}
which is a stable and positive fixed point of the mean-field dynamics. When $\eta=0$ we recover Eq.~(\ref{eq2levelfixedp}) as expected. In the absence of spontaneous excitations ($\Gamma_\text{spon} = 0$), the stable stationary solution is
\begin{equation}
n_\text{mf} = \left. \begin{cases} 
      \displaystyle 0 & \Gamma_\text{fac}\leq \frac{ (\kappa + \eta)}{1 + \frac{\eta}{\kappa^\prime}}\\
      \displaystyle\frac{\Gamma_\text{fac}(1 + \frac{\eta}{\kappa^\prime})-(\kappa+\eta)}{\Gamma_\text{fac}\left(2 + 3 \frac{\eta}{\kappa^\prime} + \left(\frac{\eta}{\kappa^\prime}\right)^2\right)} &  \Gamma_\text{fac}> \frac{ (\kappa + \eta)}{1 + \frac{\eta}{\kappa^\prime}},\\ 
   \end{cases}
\right.\label{eq3levelsfixedp_withoutspon}
\end{equation}
which also yields the analogous two-level result, Eq.~(\ref{eq2levelsfixedp_withoutspon}), for $\eta=0$. As in the two-level situation, while Eq.~(\ref{eq3levelsfixedp_withoutspon}) reflects a continuous phase transition, Eq.~(\ref{eq3levelfixedp}) describes the crossover observed in the presence of spontaneous excitations ($\Gamma_\text{spon} > 0$). From Eq.~\eqref{eq3levelsfixedp_withoutspon} we infer that the critical value of $\Gamma_\text{fac}$ in this case shifts to $(\kappa + \eta)/(1 + \frac{\eta}{\kappa^\prime})$, which is smaller (larger) than the two-level value, $\kappa$, for $\kappa^\prime < \kappa$ ($\kappa^\prime > \kappa$).

In order to illustrate the dependence of the density of excitations in the stationary state on the driving, we include the $n_\text{mf}(\Omega)$ curves for the two-level system (we choose the same parameter values used in the main text --see caption for the details), Fig. \ref{fig3levels} (a) and we plot analogous curves for the three-level system, Fig. \ref{fig3levels} (b). For the latter, we consider that the typical time scale for the departure rate to other Rydberg levels is $200\ \mu\text{s}$, $\eta = 5$ kHz, and the lifetime of the additional Rydberg level $|a\rangle$ is  $350\ \mu\text{s}$, $\kappa^\prime = 2.86$ kHz. These parameter values are in agreement with what is observed in the experiments, where the detected population of excited atoms is $n_\text{mf} + m_\text{mf} = (1 + \eta/\kappa^\prime)\, n_\text{mf} = 2.75\,  n_\text{mf}$.  The existence of blackbody radiation that results in a transition of atoms to nearby Rydberg states only modifies the quantitative details of the picture described in the main text.

\section{Probability of occurrence of occupied sublattices in a chain in the presence of quenched disorder}

In this section we substantiate the remarks that appear in the main text about the effects of quenched disorder on the dynamics. We shall take a one-dimensional configuration for simplicity. We first divide the space in which the atoms move in small enough cells (of length $a$) so that double occupancy is extremely unlikely. Each of these cells will constitute a site of an idealized lattice, which might or might not be filled with an atom. To highlight the importance of atomic motion, we show that in its absence the behavior of the system is strongly affected by finite-size effects. We further assume that the subdivision of the space is tuned in such a way that the facilitation shell produced by an excitation on the $i$-th site approximately covers only two cells at a distance $ka$ away, i.e., the two cells at positions $i \pm k$. Because of this, an excitation in $i$ can facilitate a nearby atom if either $i+k$ or $i-k$ is filled with an atom (or both are). Generalizing this picture, excitation clusters can only grow to extensive sizes if the lattice includes $k$-periodic sequences of filled sites.

\begin{figure}[h!]
\includegraphics[scale=0.14]{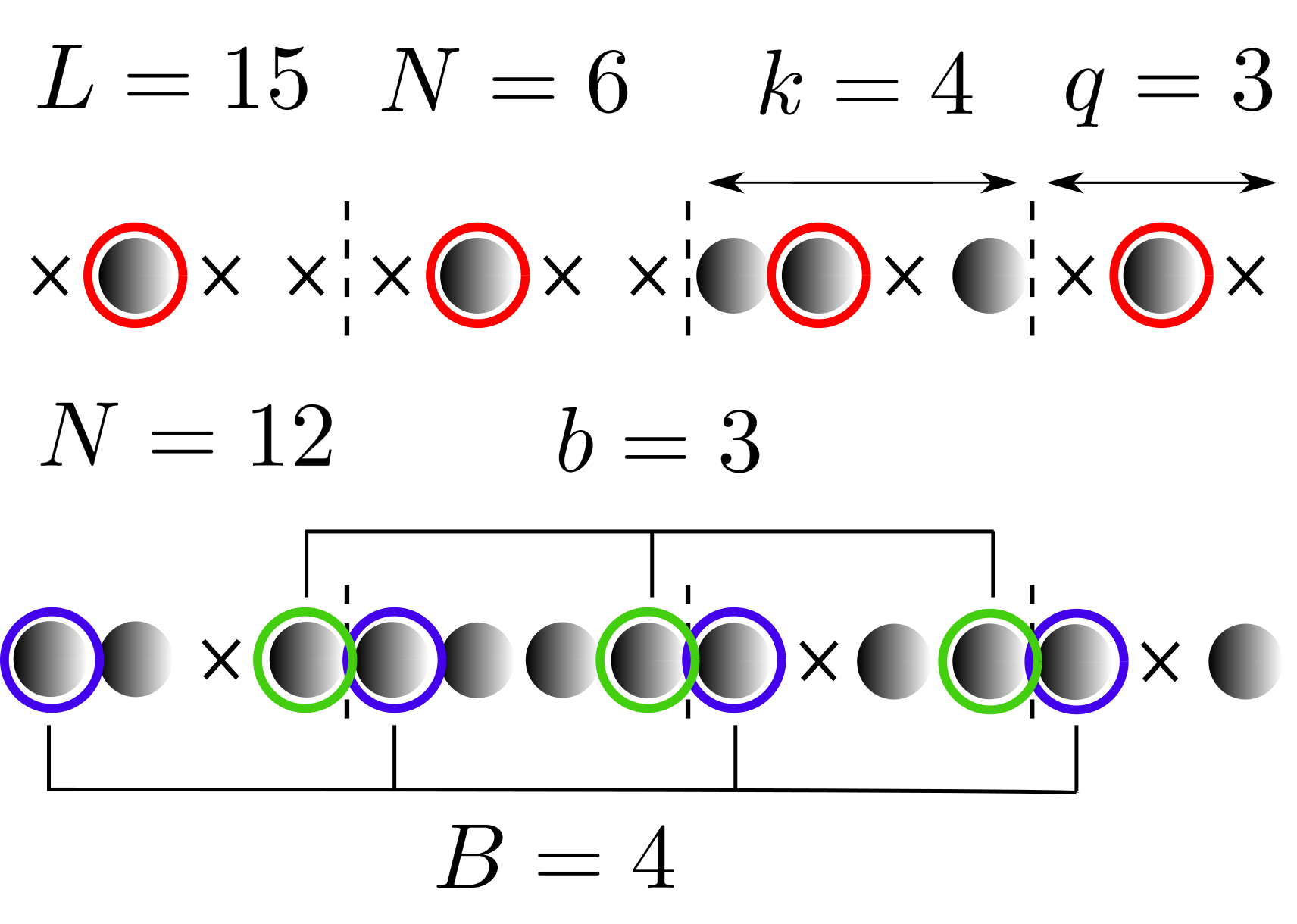}
\caption{ {\sf \bf Configurations of a chain of $L$ sites and $N<L$ atoms including a sublattice of atoms with period $k$.}
Two chains of $L=15$ sites for $k=4$. Discs correspond to filled sites (i.e. sites where an atom is present). In the chain above, $N=6$ and there is one occupied sublattice (encircled discs). As $q = L \mod k = 3 \neq 0$, sublattices can be of two different sizes. In the chain below,  which is denser -- $N=12$ -- two occupied sublattices (one comprising $B=4$ sites, the other $b=3$ sites) coexist.}\label{fig2SM}
\end{figure}

In the following, we show that the probability of occurrence of such sequences is exponentially suppressed in a static disordered system. To do so, we consider a chain of $L$ sites and $N$ atoms ($N<L$) with lattice spacing $a = r_\text{fac}/k$.
The setting is illustrated in Fig. \ref{fig2SM}, where we show two chains of $L=15$, with $N=6$ (above) and $12$ (below) atoms (crosses correspond to empty sites, discs to filled sites). In the $N=6$ case we see an occupied sublattice of periodicity $k=4$ (see encircled discs). However, sublattices are not all necessarily of the same size: if  $q \equiv L\! \mod k \neq 0$ they can comprise $B \equiv \lceil \frac{L}{k} \rceil$ or $b \equiv \lfloor \frac{L}{k} \rfloor$ sites. Indeed, in Fig. \ref{fig2SM}, where $q=3$, the $N=12$ chain shows the coexistence of sublattices of size $B = 4$ and $b = 3$.

Generally speaking, in a chain of $L$ sites and $N$ atoms ($N<L$), there can be $q = L\! \mod k$ sublattices of period $k$ comprising $B = \lceil \frac{L}{k} \rceil$  sites, and $k-q$ sublattices of the same period comprising $b = \lfloor \frac{L}{k} \rfloor$ sites.  One could thus naively expect that the number of configurations comprising at least a sublattice of period $k$ is
\begin{equation}
q {L-B \choose N - B} + (k-q) {L-b \choose N - b},
\label{marcuzzi1}
\end{equation}
where the binomials count the number of ways of filling the remaining L-B(b) sites with N-B(b) atoms. However, Eq. (\ref{marcuzzi1}) overcounts the number of configurations including at least one occupied sublattice, as it does not account for the possibility of having configurations that simultaneously include more than one -- see Fig. \ref{fig2SM} where the $N=12$ configuration would be effectively counted twice. The following term, which accounts for the number of configurations in which two or more sublattices of period $k$ are present in the system, has to be subtracted in order to correct for this form of overcounting. This term reads
\begin{equation}
{\!q \choose \! 2}\! {\!L\!-\!2 B\! \choose \!N\! -\! 2 B\!}\! +\! {\!k\! -\! q \choose 2\!}\! {\!L\!-\!2 b \choose N\! -\! 2 b} + q(k\!-\!q)\!{\!L\!-\!B\!- b \choose \! N\! -\! B\! -\!  b}. 
\label{marcuzzi2}
\end{equation}
But we again face a similar problem, as Eq.~(\ref{marcuzzi2}) overcounts all contributions in which there are more than two occupied sublattices. We have to remove from this term (and hence add back again into our count) the contribution from configurations including three or more sublattices. By iterating this procedure we arrive at the following expression for the probability of having at least one occupied sublattice of period $k$ in a chain of $L$ sites and $N$ atoms 
\begin{equation}
p(k | L,N)\! =\! \frac{\displaystyle\sum_{n=0}^q \sum_{m=0}^{k-q}\! {\!q\! \choose\!n\!}\! {\!k\!-\!q\! \choose \!m\!}\! {\!L\! -\! n B\! -\! m b\! \choose\! N\! -\! n B\! -\! m b\!}\! (-1)^{n\!+\!m\!+\!1\!}}{\displaystyle {L \choose N}}+1.
\label{marcuzziformula}
\end{equation}
The last addend ``$+1$'' cancels out exactly the $n = m =0$ term in the sum, which was added to make the notation more compact.

\begin{figure}[h!]
\hspace{0.2cm}\includegraphics[scale=0.3]{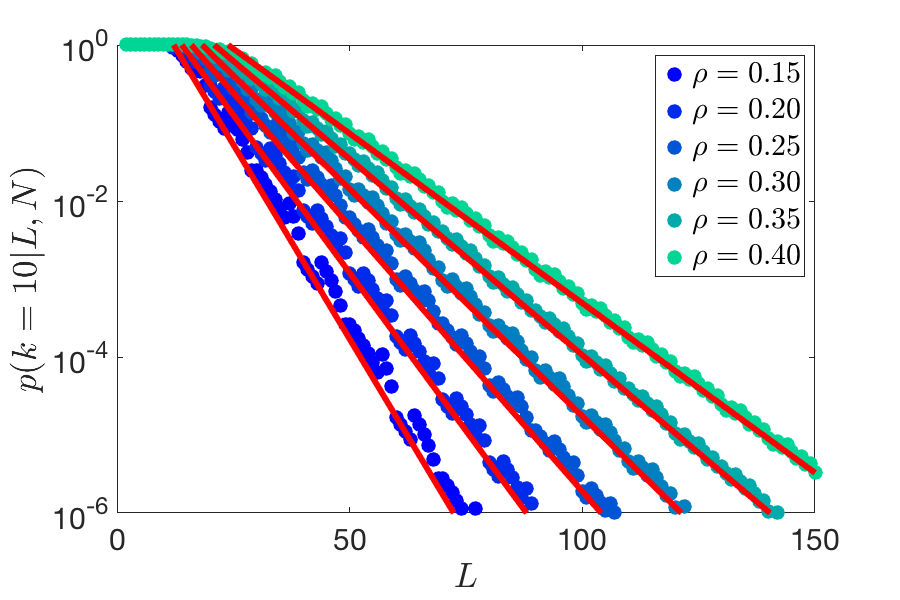}
\caption{ {\sf \bf Probability of occurrence on an occupied sublattice $p(k=10 |L,\rho L)$ as a function of $L$, and exponential approximation for large $L$.}
Probability of occurrence of an occupied sublattice of $k=10$ as function of the chain length $L$ for different densities $\rho$ ($N = \lfloor \rho L \rfloor$). Data points correspond to formulat Eq. (\ref{marcuzziformula}), with different colors distinguishing between different values of the density $\rho$. Red lines correspond to the approximation for large $L$ given in Eq. (\ref{eqnexpdec}.) for $n=1$.}\label{fig3SM}
\end{figure}

In Fig. \ref{fig3SM} we show $p(k=10|L,N = \rho L)$ as a function of $L$ for different densities $\rho$. For large systems, the probability of finding at least one sublattice of period $k=10$ decays approximately exponentially with $L$. Indeed, under such conditions Eq. (\ref{marcuzziformula}) can be approximated by $p(k|L,\rho) \propto \exp[-L/\xi(k,\rho)]$, an expression that reflects an exponential decay over a characteristic (reduced) length $\xi(k,\rho)$ that grows with $\rho$, as we now explain.

We now study the asymptotic behavior of $p(k|L,N)$ for large $L$ with a fixed density $\rho=N/L$. As $L$ grows, the difference between $B = \lceil \frac{L}{k} \rceil$ and $b = \lfloor \frac{L}{k} \rfloor$ in Eq.~(\ref{marcuzziformula}) becomes less and less relevant, and therefore to calculate the properties of $p(k|L,N)$ for long $L$ we will use the simplified version $p(k |L,N)\! \approx\! \sum_{n=1}^{k}\! {k \choose n}\! {L -n B \choose N - n B} (-1)^{n\!+\!1\!}/ {L \choose N}.$ In considering the asymptotic behavior for $L \to \infty$, we keep the density $\rho = N/L$ and $k$ fixed. The $n$-th term in the sum thus reads
\begin{equation}
	q_n(k | L,\rho) = \displaystyle{k \choose n}  \displaystyle{L(1-nk^{-1})  \choose L(\rho - nk^{-1})} / \displaystyle{ L \choose L\rho},
\end{equation}
as long as $n \leq k\rho$. For $n > k\rho$ the binomials vanish and the terms can be neglected. For the specific case $n = k\rho$ (the last non-trivial addend in the sum) one has
\begin{equation}
	q_{k\rho} (k | L, \rho) = \displaystyle{k \choose k\rho} / \displaystyle{ L \choose L\rho}.
\end{equation}
 Using Stirling's approximation ($a! \approx \sqrt{2\pi a} (a/e)^a$ for large $a$) we can see that the denominator diverges as
\begin{equation}
	\displaystyle{ L \choose L\rho} \approx \sqrt{\frac{1}{2\pi L \rho (1-\rho)} } e^{-L (\rho \log \rho  + (1-\rho) \log (1-\rho))} = \sqrt{\frac{1}{2\pi L \rho (1-\rho)} } e^{L H(A)}.
\end{equation}
with $H(A) = - \rho \log \rho  - (1-\rho) \log (1-\rho)$ the Shannon entropy for the Bernoulli trial of finding an atom at a random position. This has two outcomes $A = 1$ (atom found) and $A=0$ (not found), with probabilities $\mathbb{P}(A=1) = \rho$ and $\mathbb{P} (A = 0) = 1-\rho$, respectively.

For all the remaining terms $n < k\rho$ we can apply Stirling's formula to the numerator as well, finding
%
%
%
%
\begin{equation}
	q_n(k | L,\rho) \approx \displaystyle {k \choose n} \sqrt{\frac{\rho (1-nk^{-1})}{\rho - nk^{-1}}} e^{L f_n(k,\rho)},
\label{eqnexpdec}
\end{equation}
where $f_n(k,\rho) =  (1 - nk^{-1}) \log{(1 - nk^{-1})} + \rho \log{\rho} - (\rho - nk^{-1}) \log{(\rho - nk^{-1})}$. This exponent can be given the following interpretation: by adding and subtracting a few terms, it can be recast as
\begin{equation}
\begin{split}
	f_n(k,\rho)& = \left[ (1-nk^{-1}) \log(1-nk^{-1}) + (nk^{-1}) \log(nk^{-1})  \right] + \left[ (1-\rho) \log(1-\rho) + (\rho) \log(\rho)  \right] \\ 
	&- \left[  (nk^{-1}) \log(nk^{-1}) + (\rho - nk^{-1}) \log (\rho - nk^{-1}) + (1-\rho) \log(1-\rho)     \right] \\
	&= -H(P) - H(A) + H(A,P) = -I(A,P),
	\label{eq:f_n}
\end{split}
\end{equation}
where $I(A,P)$ is the mutual information between the two random variables $A$ defined above and $P$, corresponding to finding a site belonging to one of the considered periodic sequences by choosing one randomly. Clearly, this is also a Bernoulli trial with probabilities
\begin{equation}
	\mathbb{P}(P = 1) = nk^{-1} \quad \text{and} \quad \mathbb{P}(P=0) = 1 - nk^{-1}.
\end{equation}
However, since by assumptions all of these periodic sites are filled with atoms, $A$ and $P$ are correlated; in particular,
\begin{equation}
\begin{split}
	&\mathbb{P} (A = 1, P = 1) = nk^{-1} \\
	&\mathbb{P} (A = 1, P = 0) = \rho - nk^{-1} \\
	&\mathbb{P} (A = 0, P = 1) = 0 \\
	&\mathbb{P} (A = 0, P = 0) = 1 - \rho.
\end{split}
\end{equation}
which indeed proves that the mutual information between $A$ and $P$ is non-vanishing and corresponds to $-f_n(k,\rho)$ in Eq.~\eqref{eq:f_n}.
The mutual information increases with $n$: indeed, the larger the number of periodic sequences considered, the higher the probability of being in one by picking a random \emph{filled} site. The limiting case is the one discussed at the beginning, i.e., $n = k\rho$, in which every filled site belongs to a periodic sequence and the outcome of $A$ completely determines the outcome of $P$ (actually $A = P$). Therefore, $f_n$ is a negative and decreasing function of $n$ (as one could also check by formally taking the derivative $\partial_n$). The leading term is $n=1$ --- $n=0$ having been subtracted --- while all the other ones are exponentially suppressed with respect to it and can be neglected as a first approximation. Note that for the Stirling approximation to hold for all the terms in the sum, and thereby for the leading role of the first term to emerge as pointed out above, we have to require $L \gg k, \rho^{-1}, (1-\rho)^{-1}$.

The validity of this approximation for reasonable parameter choices and lengths $L$ only a few times larger than $k$ is shown in Fig. \ref{fig3SM} (see the red continuous lines), where $q_1(k|L,\rho) = k \sqrt{\rho(1-k^{-1})/(\rho-k^{-1})} \exp[-L/\xi(k,\rho)]$ with characteristic (reduced) length $\xi(k,\rho)=-1/f_1(k,\rho)>0$ has been used (without any form of fit).

\section{Parameters of the 1D model with motion}

In the 1D model with atomic hopping that we explore to analyze the role of atomic motion, there are two kinds of parameters. Those that are related to the excitation dynamics (namely, $\Omega$, $\Delta$, $\gamma$ and $\kappa$) can be taken as equal to the best estimates we have for the experimental values, as done also in the case of the mean-field model studied in the main text.  Other parameters (related to structural features and to motion) such as the length of the chain $L$, the density $\rho=N/L$ and the hopping rate $\lambda$ have to be inferred in a more indirect way. The criterion used for choosing a value of $L$ is straightforward and has been provided in the main text. In this section we focus on $\rho$ and $\lambda$, for which we provide ranges of values that are experimentally plausible.

In order to estimate the relevant range of densities in the experiment, we must calculate the volume of the facilitation shell, which we take to be bounded by the radial distances at which the rates, Eq.~(\ref{rates}),  equal $\Gamma_\textrm{fac}/2$, i.e. $r_{\pm} = \left[C_6/\hbar\left(\Delta \mp \gamma \right)\right]^{1/6}$. We focus on the case where $\Delta/2\pi = 10\  \textrm{MHz}$, though the sixth-root dependence of $r_\textrm{fac}$ on the detuning guarantees that the results are only weakly dependent on small or moderate changes in $\Delta$. A reasonable definition of the facilitation shell is the intersection of the spherical shell bounded by  $r_{+} = 6.745\ \mu\textrm{m}$ and $r_{-} = 6.589\ \mu\textrm{m}$ (for this choice of $\Delta$) for an atom located at the center of the blue beam, and the cylinder of diameter $w = 8\ \mu\text{m}$ (the waist of the blue laser beam) that contains the most intense part of the laser field. Accordingly, the volume of the facilitation shell on one side of an excited atom $V_\text{fac}/2 = (2\pi/3) \left[r_+^3 - r_-^3\right](1-\sqrt{1-w^2/4 r_+^2}) =8.49 \ \mu\textrm{m}^3$. As the peak density of the MOT cloud is $\rho_\textrm{peak} = 4.5 \cdot 10^{-2}\ \mu\textrm{m}^{-3}$, at the center of the cloud the number of atoms per facilitation volume is $0.76$, which is clearly an upper bound for the expected value for the system as a whole. In the numerical exploration of our model (where there is only one site per facilitation radius) $\rho = N/L$ have to be considerably smaller than this value, and we choose to focus on the case $\rho = 0.3$.

We next address the issue of atomic mobility in the experiment (which is of a ballistic nature) and in the model (which is diffusive). Our aim is to find a range of $\lambda$ that guarantees that the time it takes for an atom to move over the typical interparticle distance at the center of the cloud $l_\text{peak} = \rho_\textrm{peak}^{-1/3} = 2.81\ \mu\textrm{m}$ is similar in both situations.  In the experiment, the cloud is at $T=150\, \mu \text{K}$, and therefore we can assume that the distribution of atomic velocities follows a Maxwell-Boltzmann distribution for that temperture. As the atomic mass of Rb is $m = 87\, \text{amu} = 144.470 \cdot 10^{-27}$ kg, the average speed is $\langle v \rangle = \sqrt{8\, k\, T/\pi m} = 0.191\ \text{m s}^{-1}$. As a result it takes $\tau = l_\text{peak} /\langle v \rangle= 14.71\ \mu\textrm{s}$ to move an interparticle distance. On the other hand, diffusion in one dimensional lattices with single occupancy is characterized by a crossover from a ballistic short-time behavior to a long-time regime in which the mean-square displacement grows in time as $t^{1/2}$ \cite{vanbeijeren1983SM}. We numerically studied the diffusion problem via Monte Carlo simulations for $L=1550$, and lattice spacing $r_\textrm{fac}/10$ ($k=10$), taking as facilitation distance $r_\textrm{fac}$ the value corresponding to $\Delta/2\pi = 10\  \textrm{MHz}$ in the experiment. For  densities $\rho=0.3$ a mobility  $\lambda = 3.0$ MHz is required for the mean-square displacement to reach $l_\text{peak}^2$ in a time $\tau$. This estimate is based on $1000$ realizations of the diffusion problem, and is a lower bound for $\lambda$ in the sense that we are considering the time needed to traverse the distance $l_\text{peak}$, which is representative only of the system at the center of the cloud. If we consider longer distances, the difference between the ballistic and the diffusive motion is expected to grow, thus resulting in the need for an increased mobility in the diffusive model in order to match the experimental times $\tau$. For example, if we consider the time needed for atoms to move across a distance of $2\, l_\text{peak}$, the mobility for $\rho=0.3$ should be increased to $\lambda = 5.4$ MHz. This gives us at least some indications of the orders of magnitude of the mobility that are relevant. In the main text, we explore the model with $\lambda$ ranging from $0$ to $20$ MHz.

\end{document}